\documentclass[aps,prxquantum,reprint,superscriptaddress,nofootinbib,longbibliography]{revtex4-2}

\usepackage{tikz}
\usetikzlibrary{arrows.meta,positioning}
\usepackage{enumitem}
\usepackage{booktabs}
\usepackage{multirow}
\usepackage{amsmath}
\usepackage{enumitem}
\usepackage{graphicx}
\usepackage{xcolor}
\usepackage{url}
\newcommand{\data}[1]{\par\medskip\noindent\textbf{Data availability.}~#1\par}

\begin{document}

\title{A hardware-safety-gated system for LLM-written native ARTIQ control code on a trapped-ion platform}

\author{Duanyang Wang}
\email{duanyang.wang@duke.edu}
\affiliation{Duke Quantum Center, Duke University, Durham, North Carolina, 27701}
\affiliation{Department of Physics, Duke University, Durham, North Carolina, 27708}

\author{Lu Qi}
\affiliation{Duke Quantum Center, Duke University, Durham, North Carolina, 27701}
\affiliation{Department of Electrical and Computer Engineering, Duke University, Durham, North Carolina, 27708}

\author{Yuanheng Xie}
\affiliation{Duke Quantum Center, Duke University, Durham, North Carolina, 27701}
\affiliation{Department of Physics, Duke University, Durham, North Carolina, 27708}

\author{Norbert M. Linke}
\affiliation{Duke Quantum Center, Duke University, Durham, North Carolina, 27701}
\affiliation{Department of Physics, Duke University, Durham, North Carolina, 27708}
\affiliation{Joint Quantum Institute, University of Maryland, College Park, Maryland, 20742}
\affiliation{Department of Physics, University of Maryland, College Park, Maryland, 20742}
\affiliation{National Quantum Laboratory (QLab), University of Maryland, College Park, Maryland, 20742}

\author{Kenneth R. Brown}
\email{kenneth.r.brown@duke.edu}
\affiliation{Duke Quantum Center, Duke University, Durham, North Carolina, 27701}
\affiliation{Department of Physics, Duke University, Durham, North Carolina, 27708}
\affiliation{Department of Electrical and Computer Engineering, Duke University, Durham, North Carolina, 27708}
\affiliation{Department of Chemistry, Duke University, Durham, North Carolina, 27708}

\date{\today}

\begin{abstract}

Large-language-model (LLM) agents can write and run experimental control code. This allows laboratory work to be conducted autonomously, as in a self-driving laboratory~\cite{seifrid2022,abolhasani2023}. However, this autonomy raises a safety problem that prior work has not addressed. Unchecked code can damage the apparatus, and there is no formal, per-operation boundary between human authorization/supervision, and agent decisions.
We present a control system that places an LLM agent in the loop of a trapped-ion experiment while enforcing such a boundary. 
The agent controls the existing Advanced Real-Time Infrastructure for Quantum physics (ARTIQ) stack through tools provided by a Model Context Protocol (MCP) server. No tool call reaches the hardware unless it carries an authorization token bound to its exact contents. 
Tokens are issued in one of two ways: automatically, by running the agent's proposed script in an isolated hardware simulation (\texttt{dax.sim}~\cite{riesebos2022}) and checking every operation against preset per-device bounds, or manually by a human operator for sensitive actions.
Within this boundary the agent develops its own experiments, rather than only calling pre-built routines. 
We deploy the system on a co-trapped $^{40}$Ca$^{+}/^{40}$CaOH$^{+}$ crystal, where the agent autonomously builds a full calibration stack and, with targeted operator guidance, closes a cross-instrument magnetic-field-stabilization loop. On a separate, independent $^{171}$Yb$^{+}$ platform, we then confirm interface-level portability by running the full control stack against the live hardware.
We systematically test token-authorization mechanism with adversarial scripts that attempt to bypass it, mapping the precise boundary of its protection and prioritizing where to strengthen it next.
Analyzing where the agent still requires human guidance, we find that its limits lie in metacognitive control, namely recognizing when a problem must be
re-framed, rather than in domain knowledge.

\end{abstract}


\maketitle

\section{Background and motivation}

Quantum physics experiments across platforms share a common operational loop: the hardware must be continuously calibrated, data collected and analyzed, and parameters adjusted between runs. In trapped-ion and neutral-atom laboratories, control systems such as ARTIQ/Sinara~\cite{kasprowicz2020} have addressed the real-time layer by providing precise nanosecond timing, device abstraction, and experiment scheduling. But the experiment loop itself remains largely manual, requiring expert judgment at each iteration to interpret measurements and decide what to change next. Automating this loop not only decouples throughput from personnel availability, but also enables systematic exploration of vastly expanded parameter regimes, with the longer-term prospect of discovering hints of new physics~\cite{zhang2026}.

Machine-learning and optimization methods have been applied across platforms to automate parts of this loop:
learned models navigate charge-stability diagrams to tune gate-defined quantum dots~\cite{moon2020,nguyen2021,schuff2026,zwolak2023}, dependency-graph systems schedule and execute superconducting-qubit calibration~\cite{kelly2018,wittler2021}, 
and Bayesian optimization and reinforcement learning have been applied to closed-loop tuning of gate parameters in trapped-ion systems~\cite{niu2019,baum2021,gerster2022}.
These methods share a structural limitation: the experimental procedure is designed and coded entirely by researchers; the algorithm only searches within the parameter space of that fixed procedure. The dependency graph of Kelly et al.~\cite{kelly2018} propagates recalibration through pre-written nodes without generating code; the Bayesian optimizer of Gerster et al.~\cite{gerster2022} navigates the parameter space of a human-designed pulse sequence without altering its structure. The burden shifts from running experiments manually to coding and maintaining the routines that automated systems rely on.

LLMs can generate executable code from natural-language descriptions and interpret unstructured outputs such as error messages and logs, capabilities that map onto the requirements of iterative experimental control. In chemistry and materials science, LLM agents have translated natural-language goals into procedures, written and executed control code, and refined strategies based on measured outcomes~\cite{boiko2023,mbran2024,lu2024,gottweis2025}. In quantum hardware, LLM agents have translated goals into calibration routines for a superconducting processor~\cite{cao2025}, assisted qubit measurement~\cite{li2026} and quantum-dot timing calibration~\cite{rao2025}, and extracted insights from calibration data~\cite{cao2026}. These quantum-hardware demonstrations, however, still rely on fixed libraries of human-written routines rather than generating new control code, and several operate only in cloud or simulation environments. An agent that generates its own control code and executes it on real quantum hardware while iterating on measured results would address both limitations, but introduces two open gaps. First, no hardware-protection layer has been formalized for this setting. Safety frameworks from autonomous chemical and materials laboratories~\cite{leong2025,zhang2026,tang2025} address failure modes that do not transfer directly to quantum hardware, and the most closely related quantum demonstration defers safety to future work~\cite{cao2025}. Second, no prior system provides a machine-checkable criterion for vetting individual hardware operations generated by an agent; existing work describes human oversight only in qualitative terms~\cite{lazaros2026}.

In this work, we address both gaps by deploying an LLM-based closed-loop automation system on real trapped-ion experiments operated through the ARTIQ/Sinara stack~\cite{kasprowicz2020}. The system comprises two main components. The first, a \textit{safety-filter}, enforces a per-call authorization gate to scripts generated by LLM agents. The second, \textit{artiq-mcp}, exposes the ARTIQ master to the agent as tools through the Model Context Protocol (MCP)~\cite{anthropic2025}, building on the development of tool use in language models~\cite{schick2023,yao2023,patil2024,xi2025}. We evaluate the framework's generality and structural robustness by deploying it across two distinct physical setups. First, on a co-trapped $^{40}$Ca$^+$/$^{40}$CaOH$^{+}$ system at Duke University~\cite{qi2023,reed2024,qi2024}, we validate the agent's capacity to autonomously synthesize complex quantum control logic from scratch and drive heterogeneous, cross-platform hardware for active magnetic-field compensation. Second, to test ecosystem portability, we perform an interface-level validation of the identical safety-gated stack on an independent ARTIQ-controlled $^{171}$Yb$^+$ system at the University of Maryland. This integration test on the live control hardware also serves as the baseline for an extensive red-team campaign of the safety filter, by which we mean deliberately attacking our own deployed system with adversarial scripts to find the boundary of its protection. The campaign comprises 1932 bench scripts and approximately 250 harness-level tests; it maps the empirical boundary of the filter's protection, quantifies the structural separation between evasion code and the code an agent writes during normal calibration, and sets out a layered plan for hardening the filter. The remainder of the paper presents the system architecture (Sec.~\ref{sec:architecture}), the experimental program (Sec.~\ref{sec: exp}), and a discussion (Sec.~\ref{sec:discussion}).

\begin{figure*}[tb]
\centering
\usetikzlibrary{arrows.meta, positioning, calc}

\begin{tikzpicture}[
  font=\sffamily,
  >={Stealth[round]},
  line cap=round,
  every path/.style={line width=0.7pt},
  box/.style={draw=#1!60!black, fill=#1!12, rounded corners=3pt,
              align=center, font=\sffamily\small, inner sep=5pt,
              minimum height=1.0cm, minimum width=2.1cm,
              line width=0.8pt},
  el/.style={font=\sffamily\scriptsize, text=black!70, fill=white,
             inner sep=1.5pt}
]
\node[box=blue]                      (agent)  at (0,0)    {AI agent};
\node[box=orange]                    (filter) at (3.8,0)  {safety\\filter};
\node[box=violet]                    (mcp)    at (7.6,0)  {\textit{artiq-mcp}};
\node[box=gray, minimum width=2.9cm] (master) at (11.4,0) {ARTIQ master\\\& hardware};
\node[box=teal]                      (human)  at (0,2.4)  {Human\\operator};
\node[box=orange]                    (sim)    at (3.8,2.4){\texttt{dax.sim}\\review};

\draw[->] (agent)  -- (filter) node[el,above=1pt,midway]{tool call};
\draw[->] (filter) -- (mcp)    node[el,above=1pt,midway]{approved};
\draw[->] (mcp)    -- (master) node[el,above=1pt,midway]{\texttt{sipyco}};

\draw[->]  (human)  -- (agent) node[el,right=1pt,midway]{goal};
\draw[<->] (filter) -- (sim)   node[el,right=1pt,midway]{simulate \& check};
\draw[->,dashed] (human.south east) -- (filter.north west)
      node[el,midway,sloped,above=0pt]{approve sensitive ops};

\draw[->, rounded corners=6pt, blue!50!black]
      (master.south) -- ++(0,-1.0) -| (agent.south)
      node[el, text=blue!50!black, pos=0.32]
      {experiment results \& logs (agent observes and iterates)};
\end{tikzpicture}
\caption{Overview of the closed-loop control path. The human operator supplies a high-level goal and approves only sensitive operations, which issue a human token. The AI agent plans, writes, and submits scripts as tool calls to the \textit{safety-filter}, which authorizes each call before forwarding it to \textit{artiq-mcp}. A call that runs an agent-written script is authorized by executing the script in an isolated \texttt{dax.sim} simulation and checking the resulting hardware trace against preset bounds. \textit{artiq-mcp} drives the ARTIQ master, which executes on the trap hardware, and the resulting datasets and logs flow back to the agent, which observes the outcome and iterates until the goal is met. The agent never communicates with \textit{artiq-mcp}, the ARTIQ master, or the hardware directly.}
\label{fig:overview}
\end{figure*}
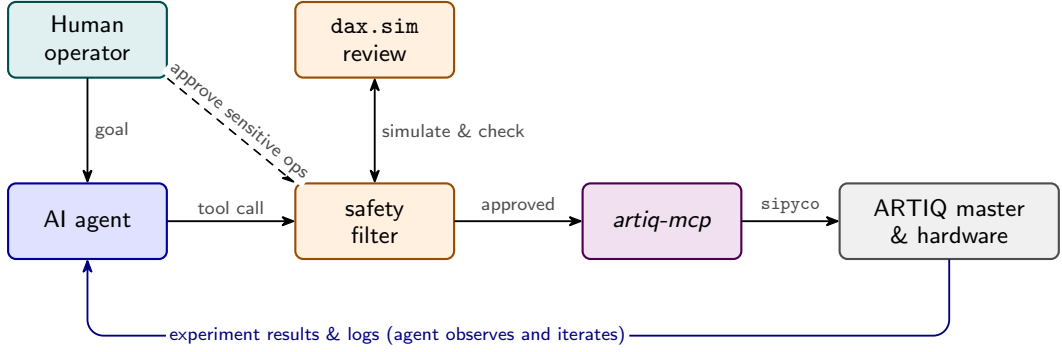

\section{System architecture}
\label{sec:architecture}

The system places an LLM agent in the control path of the laboratory apparatus through two interposed software components, such that the agent never communicates with the hardware directly (Figure~\ref{fig:overview}). Human operators provide high-level goals in natural language; the agent designs the experiment and generates the corresponding ARTIQ Python scripts, which are submitted to the hardware through MCP tool calls. Every call is received first by the \textit{safety-filter}, an MCP proxy that enforces a per-call authorization gate (Sec.~\ref{sec:filter}) and forwards only approved calls to \textit{artiq-mcp}, a host-side MCP server that exposes ARTIQ master functions as agent-invocable tools by communicating over ARTIQ's native \texttt{sipyco} interfaces. The server submits each approved call to the ARTIQ master for execution on the Sinara core device and peripheral trap hardware, then returns the resulting datasets and logs to the agent. The agent observes the outcome, diagnoses any failure, revises its script if necessary, and iterates until the stated goal is met.

Given this level of autonomy, the system is designed around a specific threat model in which the agent does not act maliciously, but may generate unsafe hardware commands due to reasoning errors or by processing adversarial content (e.g., malformed free text from experiment logs or error messages). Consequently, the protection mechanisms do not rely on the agent's cooperation. In Sec.~\ref{sec:redteam}, we deliberately relax this assumption and adopt an adversarial agent. This is not to claim robustness against determined attackers, but rather to empirically map the boundaries of the filter's protection and identify necessary future defense layers. The \textit{safety-filter}'s guarantees are strictly limited to deterministic hardware protection (e.g., enforcing RF power caps or valid logic levels); verifying whether an experimental sequence is sensible remains out of scope (Sec.~\ref{sec:filter}).

\subsection{Call-level authorization and automated safety verification}
\label{sec:filter}

The \textit{safety-filter} is the single enforcement point between the agent and the rest of the control stack. 
It operates as an MCP proxy: the agent connects exclusively to the filter, and no tool call reaches \textit{artiq-mcp} or the hardware unless the filter has issued a valid authorization token for it. Each such token is bound to a single tool call with specific contents, is consumed upon execution, and cannot be used a second time, used for a different call, or spent by two concurrent calls. The mechanism is applied to every call regardless of origin, so there is no execution path by which the agent can act on the apparatus without passing the authorization gate. The gate guarantees that a token exists; the depth of verification behind that token depends on the operator's device configuration.

Tokens are issued via either automatic or human review. The automatic review pipeline proceeds in two stages. First, an Abstract Syntax Tree (AST) checker screens the agent's script by matching call targets and import statements against a denylist of host-reachable constructs (e.g., \texttt{os.system}, \texttt{subprocess}, \texttt{eval}). This check catches known misuse patterns at the syntactic level and does not guarantee completeness; its empirical boundaries are evaluated in Sec.~\ref{sec:redteam}. Second, if the AST check passes, the filter runs the script against \texttt{dax.sim}, the simulation backend of DAX (Duke ARTIQ Extensions, a modular open-source control-software framework built on ARTIQ)~\cite{riesebos2022}. The simulator substitutes software models for physical ARTIQ devices on a separate host, producing a line-by-line trace of every hardware operation the script would perform without touching the apparatus.

Each operation, including DDS frequency, amplitude, and phase settings, DAC output voltages, and TTL pulses, is checked against per-device bounds defined within the operator's configuration. Bounds are set per device instance, allowing the operator to impose tighter limits on individual instruments as needed. The filter's response depends on the verification outcomes: if any parameter on a mapped operation exceeds its bounds, the call is blocked immediately. If the script invokes a device or method that is not explicitly mapped with limits in the operator's configuration, the call is blocked and escalated to manual review, consistent with the block-by-default policy applied when the simulation cannot reach a deterministic verdict. The operator may then choose to add the device to the configuration with appropriate bounds, or issue a manual token for the specific call. Conversely, when all operations fall within bounds, the filter stores the script's content hash to bypass re-simulation for byte-identical resubmissions. This stored approval does not expire on its own, so it leaves a window in which a script verified earlier could run under bounds that were tightened after it was stored. We close this window by operator action rather than on a timer. After tightening bounds, the operator manually invalidates the outstanding stored approval, since the filter does not itself detect the change or determine which entries it affects. We avoid blanket re-simulation of every simulated script on each boundary change, which would be too costly during routine tuning. Responsibility for approval consistency therefore rests with the operator who authors the bounds (Table~\ref{tab:vulns}).

The scope of the simulation-based review is deliberately narrow. The resulting trace exposes only hardware-layer compliance, focusing on the numerical boundaries of each real-time operation and their corresponding timing sequence. It is not meant to evaluate whether the experiment it runs is sensible, because physical behaviors such as light-atom interactions lie entirely outside the scope of the instruction trace. Instead, physical modeling errors are expected to surface during the agent's closed-loop optimization, while fundamental configuration anomalies are backstopped by manual human authorization. Confining the filter responsibility to hardware protection ensures that its safety guarantees remain verifiable and auditable.

Another structural limitation stems from measurement-conditioned control flow. The simulation executes each script using a single, deterministic realization of any mid-experiment measurement. For example, the fluorescence photon counting interface is modeled to always return a dark state with a count of zero, meaning the simulated trace covers only one specific branch of a conditional code path. The automatic review pipeline can therefore only guarantee safety for scripts whose hardware operations do not depend on measurement outcomes. Consequently, the agent's autonomy is concentrated on host-level experiment scheduling and iterative parameter optimization, rather than authoring nanosecond-scale adaptive branching kernels (such as real-time feed-forward or active reset) on the core device. In practice, this constraint does not affect the calibration sequences deployed in this work, which are open-loop parameter sweeps (spectral scans, sideband cooling, Rabi flopping, and Ramsey interferometry) where all hardware operations are fixed before any measurement is read. The AST checker screens for recognizable instances of such measurement-gated control flow, such as conditionals that branch directly on a PMT return value, and rejects scripts matching these patterns. As with any syntactic check (see above), equivalent logic re-expressed through indirect constructs may evade it. Exhaustively enumerating every branch is not a general remedy. Each mid-experiment measurement is a branch point, so the number of distinct execution paths grows exponentially in the number of measurements; once a measurement also gates a loop whose iteration count depends on its own outcome, the number of branches is no longer fixed at check time, and static review cannot in general certify that such a loop terminates. The simulation therefore follows a single branch and the filter does not vouch for the others; bounding this path explosion well enough to enumerate dynamic branches safely is left to future work.

Under the human review protocol, sensitive operations excluded from automated approval, primarily those risking irreversible hardware damage, require manual authorization. For these, an operator issues tokens separately, outside the automated pipeline; unlike the single-use tokens produced by automated review, human-issued tokens may carry a usage quota allowing repeated execution. These tokens persist across sessions and can be audited and revoked at any time. If the automated review cannot reach a deterministic outcome (e.g., the simulation host is unreachable), the same block-by-default policy applies and the call is escalated to manual review.

Several deployment-level measures further harden the system. First, the agent reaches the filter only over a restricted channel configured solely to communicate with the filter; interactive shells, port forwarding, and credential delegation are disabled, confining the agent entirely to the designated tool interface. Second, the filter itself runs under a dedicated, unprivileged account and serves as the exclusive pipeline between the agent and the \textit{artiq-mcp} backend. Third, to eliminate external vectors, all review, authorization, and result handling take place on the local laboratory infrastructure without external service calls. Finally, every execution call and authorization decision is written to an append-only audit log, enabling deterministic post-hoc reconstruction of the full sequence of agent actions. The trace-based architecture also accommodates more sophisticated evaluators without modifying the authorization pathway. During the initial deployment, a separate bypass tooling path remains available for out-of-pipeline execution; the practical security implications of the full architecture, including AST checker limitations and the developer bypass, are empirically evaluated in Sec.~\ref{sec:redteam}. 
 
 \subsection{Interfacing the control system with the agent}

The \textit{artiq-mcp} component is a host-side MCP server that wraps the functions of a running ARTIQ master as callable tools. Through this interface, the agent can submit experiments, monitor execution, read and write datasets, query the device database, and retrieve logs and dashboard control events. Communication with the ARTIQ master relies on its native Remote Procedure Call (RPC), subscription, and broadcast interfaces. The server does not compile or execute real-time kernels and introduces no real-time control logic, operating purely as a host-side mediator.

In practice, the agent uses these tools in a standard workflow.  It first submits an experiment to receive a unique run identifier, then polls the run status until execution completes. Upon completion, results are extracted from the standard output ARTIQ HDF5 datasets and parsed into structured formats such as JSON for interpretation by the LLM. The agent can also inspect datasets and stream logs mid-run to diagnose anomalous executions, enabling a feedback loop to revise scripts and iteratively converge on a working calibration. High-risk operations, including core-device reboots, configuration erasures, or master terminations, feature a built-in two-step confirmation native to \textit{artiq-mcp}. In this deployment these checks are redundant with the primary authorization layer of Sec.~\ref{sec:filter}, which gates every call before it reaches \textit{artiq-mcp}; we retain them as an additional layer of defense but do not rely on them for the safety guarantees reported here.

\subsection{Cross-device closed-loop operation}
\label{sec:cross_device}

Operating exclusively through the MCP interface enables the agent to integrate instruments beyond the ARTIQ ecosystem into a unified control loop. Any instrument exposed as a standalone MCP server, such as an arbitrary waveform generator (AWG) driving a magnetic-field compensation coil, becomes accessible to the agent under the same per-call authorization once the operator adds its tools to the filter's device configuration. Because these instruments lie outside the \texttt{dax.sim} simulation, their calls are authorized exclusively through human-issued tokens rather than automated trace review. This does not require native ARTIQ-kernel support for the instrument, so a closed loop can span the ARTIQ master and asynchronous external hardware without modifying either. In Sec.~\ref{sec:sixtyhz} we demonstrate this with an active magnetic-field compensation system in which the agent couples ARTIQ-based Ramsey interferometry with an external AWG to suppress 60~Hz power-line-synchronous magnetic-field noise.

Especially hazardous hardware is withheld from the agent in one of two ways. The first is hardware the filter sees and denies. On our platform this is the oxygen gas doser (Appendix~\ref{app:platform}). Its high-voltage line and gas valve appear in the trace but are marked as denied in the device configuration, and any script that actuates these devices fails review. The second is hardware the filter cannot see. On our platform this is the atom-source oven that loads ions into the trap (Appendix~\ref{app:platform}). It is driven by a bench power supply over a serial link, outside the ARTIQ real-time path. It appears in no trace, and the agent is given no interface to it.

\section{Experimental results}
\label{sec: exp}

\begin{figure*}[tb]
\centering
\includegraphics[width=0.49\linewidth]{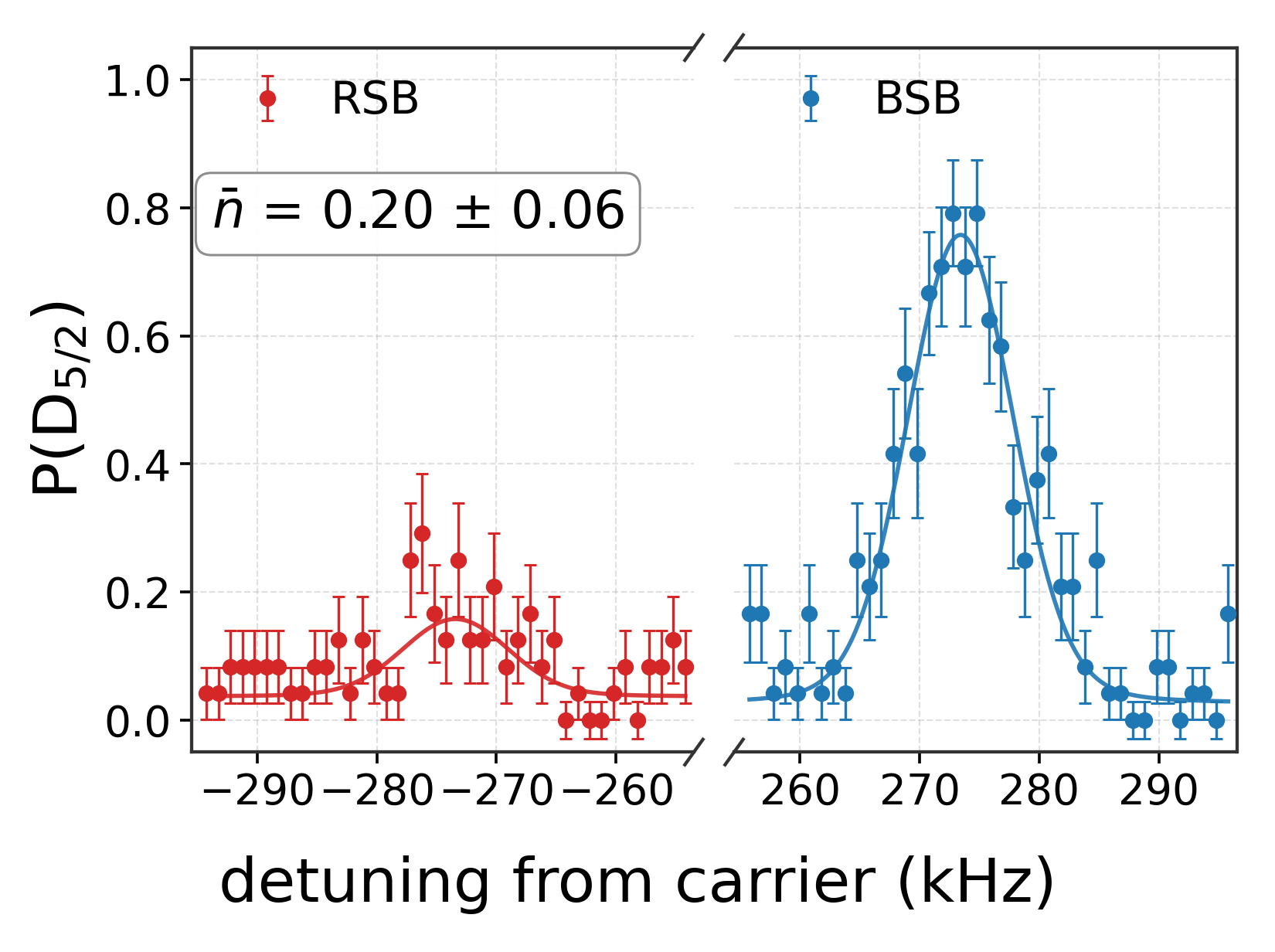}\hfill
\includegraphics[width=0.49\linewidth]{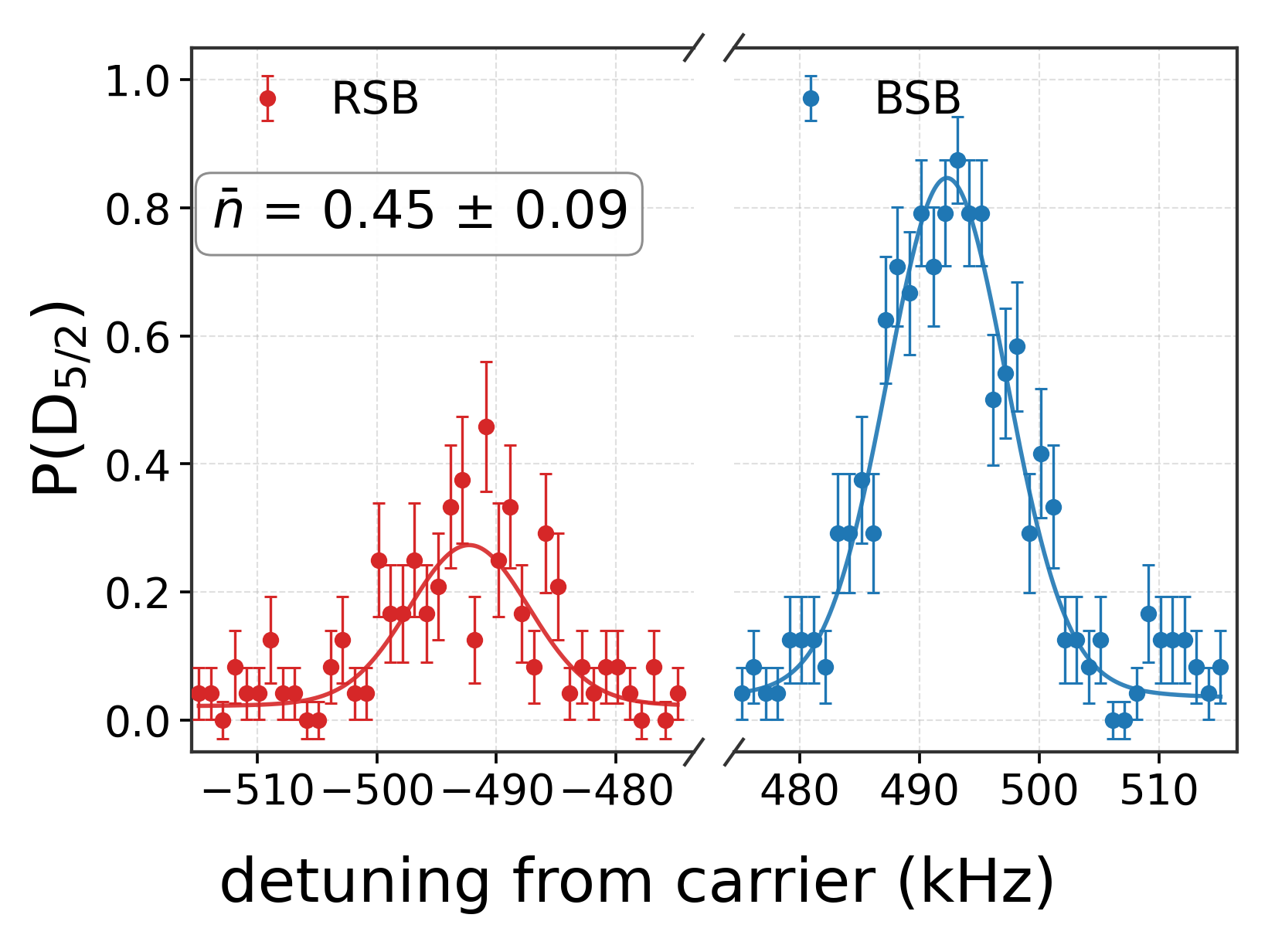}
\caption{Sideband thermometry at the endpoint of the agent-built cooling chain. Shelving probability against detuning from the 729 carrier for the axial center-of-mass mode (left) and the breathing mode (right) of the co-trapped crystal, measured after the agent's Doppler and continuous sideband-cooling sequence. Each panel shows the red-sideband (RSB) and blue-sideband (BSB) resonances; solid lines are independent Lorentzian fits, and error bars are binomial uncertainties of the shelving probability. The sideband ratios give mean occupations $\bar{n} = 0.20 \pm 0.06$ (center of mass) and $0.45 \pm 0.09$ (breathing). Each point is the shelving probability estimated from N=100 binary (bright/dark) shots, and the error bar is the binomial standard error $\sqrt{p(1-p)/N}$.}
\label{fig:sbc}
\end{figure*}

\begin{figure*}[tb]
\centering
\includegraphics[width=0.49\linewidth]{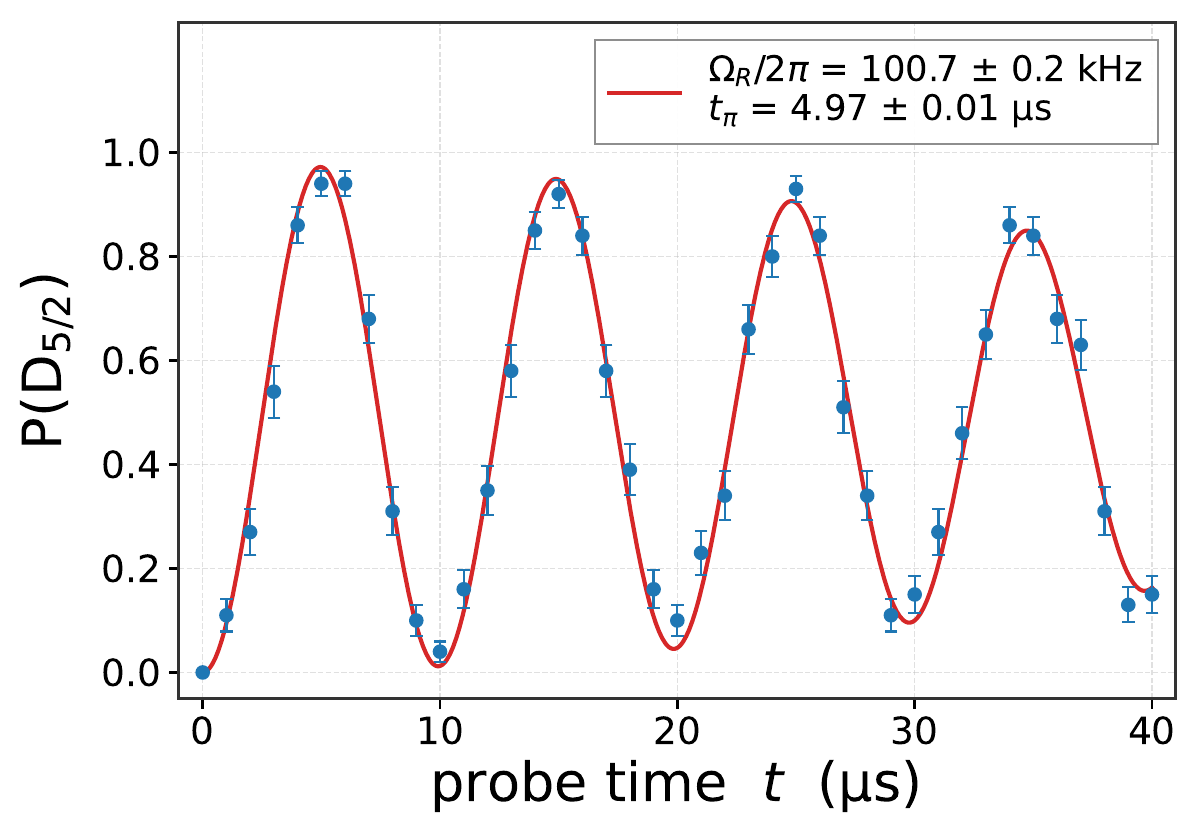}\hfill
\includegraphics[width=0.49\linewidth]{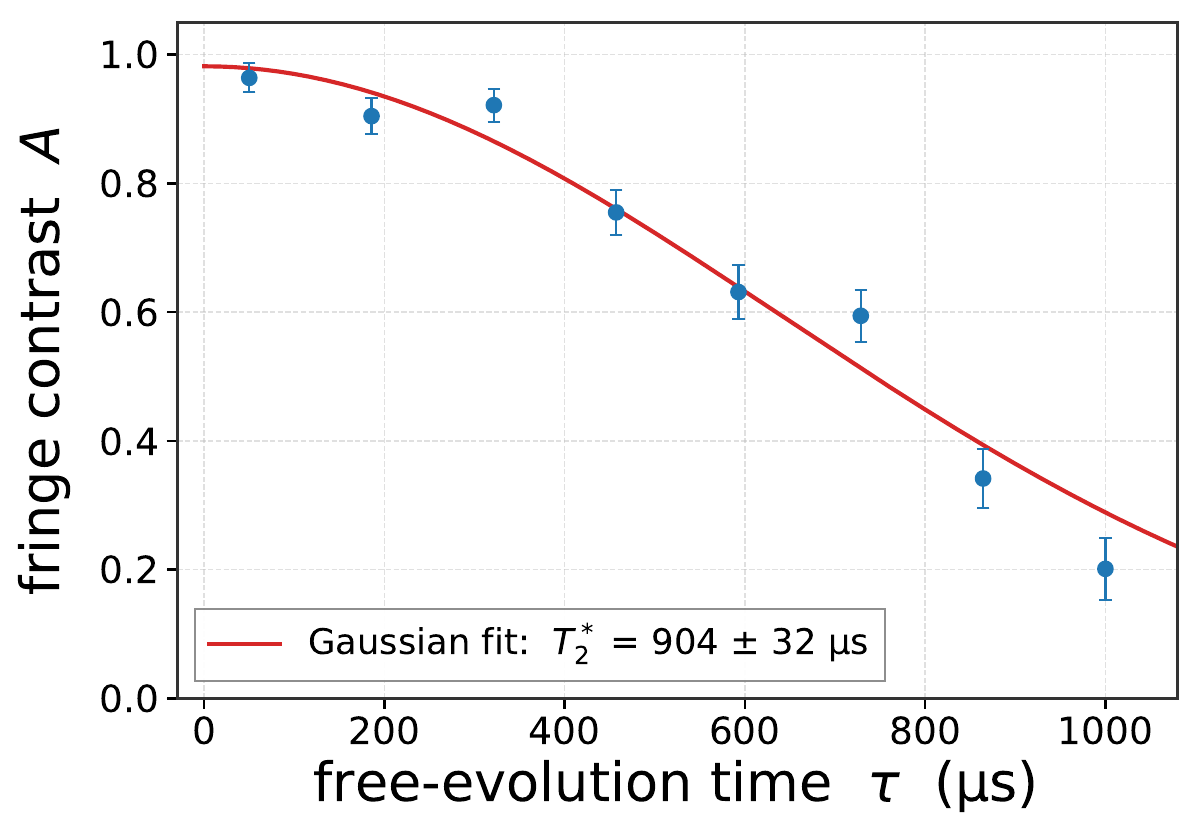}
\caption{Carrier Rabi flopping on the 729 transition (left), measured by the agent's probe sequence after cooling. Shelving probability against probe pulse duration; error bars are binomial standard error over the $N=100$ measurement sweeps. The data span four full Rabi cycles. The solid line is a fit to a damped Rabi model, giving a contrast of 0.975, a $\pi$ time of $4.97\,\mu$s, and a Gaussian decay constant of $63.8\,\mu$s. Ramsey calibration built by the agent (right). Fringe contrast against free-evolution time, extracted from phase-scanned fringes at 8 delays; error bars are binomial shot errors propagated through the fringe fits (appendix~\ref{app:uncertainty}). The solid line is a Gaussian envelope fit giving $T_2^* = 904 \pm 32\,\mu$s.}
\label{fig:rabi}
\end{figure*}

All experiments reported below were carried out by a single coordinating agent based on Claude Opus 4.8 (Anthropic), run with extended reasoning (the ``extra effort'' setting) and a one-million-token context window. Because no context window can span an entire campaign, the agent did not run as one continuous session but worked across many, carrying its goal, curated working memory, and record of progress between them in external Markdown files that it reloaded at the start of each session. To handle the breadth of each task, the agent spawned short-lived subagents for delimited subtasks such as code generation, log parsing, or fit analysis, but it held the experimental goal and remained the only agent connected to the safety filter throughout, with no separate agent in a supervisory role, the structural point we return to in Sec.~\ref{sec:capability}.

The experimental results are organized by platform and objective. Section~\ref{sec:corecal} details the core evaluation on the molecular-ion platform, where the agent autonomously constructs a full calibration stack on the co-trapped crystal, encompassing cooling, Rabi, and Ramsey sequences driven by a small number of high-level prompts. Section~\ref{sec:sixtyhz} extends this capability to a multi-MCP scenario, demonstrating how the agent closes an active magnetic-field compensation loop across heterogeneous instruments to suppress 60~Hz power-line noise. To evaluate generalizability, Section~\ref{sec:portability} reports a cross-platform deployment on a separate, independent ARTIQ system. Section~\ref{sec:redteam} then presents an adversarial boundary characterization of the safety filter.

\subsection{Agent-built calibration of a co-trapped crystal}
\label{sec:corecal}

The molecular-ion platform is a two-ion Coulomb crystal that co-traps an atomic $^{40}$Ca$^{+}$ ion with a molecular $^{40}$CaOH$^{+}$ ion in a linear Paul trap~\cite{qi2024, qi2023, reed2024}. The atomic ion carries all of the laser interactions used here. Its 397\,nm and 729\,nm transitions provide Doppler cooling, state preparation, coherent control, and shelving readout, while the molecular ion is held in the same potential and cooled sympathetically through the shared motion. The calibrations reported in this section act on the atomic ion and on the shared axial motion, so they are independent of the molecular internal state. A fuller description of the trap and the relevant atomic structure is given in Appendix~\ref{app:platform}.

The calibration stack was built in five sequential stages. Each stage was initiated by a short natural-language prompt together with the relevant hardware parameters. The agent submitted its generated ARTIQ code through the safety filter and used the resulting runtime feedback to drive the optimization loop, stopping only upon internal convergence or operator approval. Table~\ref{tab:calibration_stages} outlines these stages, detailing the specific parameters supplied at each step and the corresponding operator advice required.

\paragraph{Doppler cooling.}
The first stage was initiated with the prompt: \textit{``Develop a Doppler cooling experiment for the \textsuperscript{40}Ca\textsuperscript{+}/\textsuperscript{40}CaOH\textsuperscript{+} ion chain.''}. In addition to the hardware interface document, the operator provided the 397\,nm laser alignment geometry, the optical power as a function of AOM attenuation, and the approximate beam dimension at the ion. From these inputs the agent determined the operating AOM attenuation and wrote the cooling sequence. The module was not separately optimized at this stage.  Its adequacy was confirmed downstream, when the Rabi experiment built on top of it yielded a $\pi$ time and contrast consistent with those obtained in routine manual operation. This indicates that the agent's cooling configuration was already sufficient.

\paragraph{729\,nm spectroscopy.}
The second stage was initiated with:
\textit{``Use the 729\,nm laser to scan the frequency spectrum of the quadrupole transition.''}. The operator provided the 729\,nm laser alignment, beam polarization, the static magnetic field magnitude, the axial secular frequencies of the ion chain, and approximate values for the carrier AOM setpoint and expected Rabi frequency. The agent performed a coarse, high-power, short-pulse scan to locate the Zeeman manifold of the S\textsubscript{1/2}$\leftrightarrow$D\textsubscript{5/2} quadrupole transition, followed by a narrow, low-power, long-pulse scan to resolve and fit the carrier line. The fitted carrier frequency was written to the \texttt{ai.*} dataset namespace and read by all subsequent experiments.

\paragraph{Resolved sideband cooling.}
The third stage directly integrated the Doppler cooling and spectroscopy modules developed in the preceding phases. The operator prompt was: \textit{``Use the 729\,nm laser to cool the ion's axial motion.''}. 
The agent proposed and implemented a pulsed sideband cooling scheme in the beginning: alternating red-sideband pulses and 854\,nm quench windows, with $\sigma^{+}$ optical pumping to spin-polarize the ion before each cooling cycle. To optimize phonon removal efficiency, the agent structured the sequence into multiple progressive stages, scheduling near-$\pi$ pulses calculated for the initial thermal distribution at high $\bar{n}$, followed by longer pulses matched to the decaying $\bar{n}$. 
This cooling sequence was optimized iteratively against sideband thermometry. 
Statistical fluctuations in early stage runs prompted operator advice to measure the red and blue sidebands by scanning and fitting rather than single-point reads, increase the shot count per scan point, and adopt a reduced $\chi^{2}$ goodness-of-fit threshold as the convergence criterion, which successfully stabilized the thermometry fits. 
The initial pulsed implementation required a total duration of approximately 80\,ms. 
Subsequently the operator feedback and demand for a faster duty cycle prompted the agent to transition to a continuous sideband cooling scheme~\cite{reed2024}. 
By applying the red-sideband drive and 854\,nm quench simultaneously rather than in discrete alternating intervals, the agent rewrote the cooling module to compress the total duration to 19~ms while maintaining comparable final motional occupations.

\paragraph{Rabi flopping and Ramsey interferometry.}
Each of these two stages was initiated by a single-line prompt specifying the measurement type and the target carrier frequency. In both cases, the agent composed the experiment by calling the Doppler cooling, spectroscopy, and sideband cooling modules as prerequisites, then appended the relevant coherent-pulse sequences and shelving readouts. The Ramsey experiment re-ran the 729\,nm carrier scan at the start of each session to track slow frequency drifts before acquiring fringes. Both stages achieved successful execution without requiring operator advice. The endpoints of the completed calibration stack are shown in Figures~\ref{fig:sbc} and~\ref{fig:rabi}. After Doppler and continuous sideband cooling, red- and blue-sideband shelving scans of the two axial modes give mean occupations $\bar{n}=0.20\pm0.06$ (center-of-mass) and $0.45\pm0.09$ (breathing mode), confirming that both modes are cooled near the motional ground state (Figure~\ref{fig:sbc}). Characterization of the carrier Rabi oscillations indicates a state contrast of 0.975, a $\pi$ time of $4.97\,\mu$s, and a Gaussian decay constant of $63.8\,\mu$s (Figure~\ref{fig:rabi}, left). The Ramsey measurement, with phase-scanned fringes at eight free-evolution times between 50 and 1000\,$\mu$s, gives a coherence time of  $T_{2}^{*}=904\pm32\,\mu$s (Figure~\ref{fig:rabi}, right).

\begin{table*}[ht]
\caption{Summary of calibration stages. Each stage was initiated by
a natural-language prompt; all parameter scanning, code
generation, and convergence decisions were handled by the agent.}
\label{tab:calibration_stages}
\begin{center}
\begin{tabular}{p{2.4cm} p{4.0cm} p{3.8cm} p{3.2cm}}
\hline
\textbf{Stage} &
\textbf{Initiating prompt} &
\textbf{Parameters supplied by operator} &
\textbf{Operator advice} \\
\hline
1.\ Doppler cooling &
\textit{``Develop a Doppler cooling experiment for the
$^{40}$Ca$^{+}$/$^{40}$CaOH$^{+}$ ion chain.''} &
397\,nm alignment; power vs.\ AOM attenuation; beam position at ion &
None at this stage \\
\hline
2.\ 729\,nm spectroscopy &
\textit{``Use the 729\,nm laser to scan the quadrupole transition
spectrum.''} &
729\,nm alignment; polarization; $B$-field; axial secular
frequencies; carrier AOM setpoint $\approx$152\,MHz;
$\Omega/2\pi \approx 100$\,kHz &
None \\
\hline
3.\ Sideband cooling &
\textit{``Use the 729\,nm laser to cool the ion's axial motion.''} &
Inherited from stages 1 and 2 &
Data-quality directives (shot count, $\chi^{2}$ criterion);
request for shorter sequence \\
\hline
4.\ Rabi flopping &
\textit{``Develop Rabi flopping on the carrier at $X$\,MHz.''} &
Calibrated carrier frequency from stage 2 &
None \\
\hline
5.\ Ramsey interferometry &
\textit{``Build a Ramsey fringe measurement.''} &
Inherited from stages 1--3 &
None \\
\hline
\end{tabular}
\end{center}
\end{table*}

\subsection{Closed-loop power-line magnetic-field compensation}
\label{sec:sixtyhz}

This experiment extends the agent beyond the ARTIQ master to a multi-instrument closed loop, exercising the cross-MCP capability of Sec.~\ref{sec:cross_device}. The target was power-line-synchronous (60\,Hz) magnetic-field noise, the dominant limit on the carrier Ramsey coherence time $T_2^{*}$ of the optical qubit. While native ARTIQ hardware is optimized for deterministic digital timing, it lacks the analog infrastructure to directly synthesize the complex, multi-harmonic waveforms required for this cancellation. By integrating an external AWG as a dedicated analog actuator, the agent could simultaneously command ARTIQ-based Ramsey measurements and reparametrise the external compensation waveform within a single control loop.

\begin{figure*}[tb]
\centering
\includegraphics[width=\linewidth]{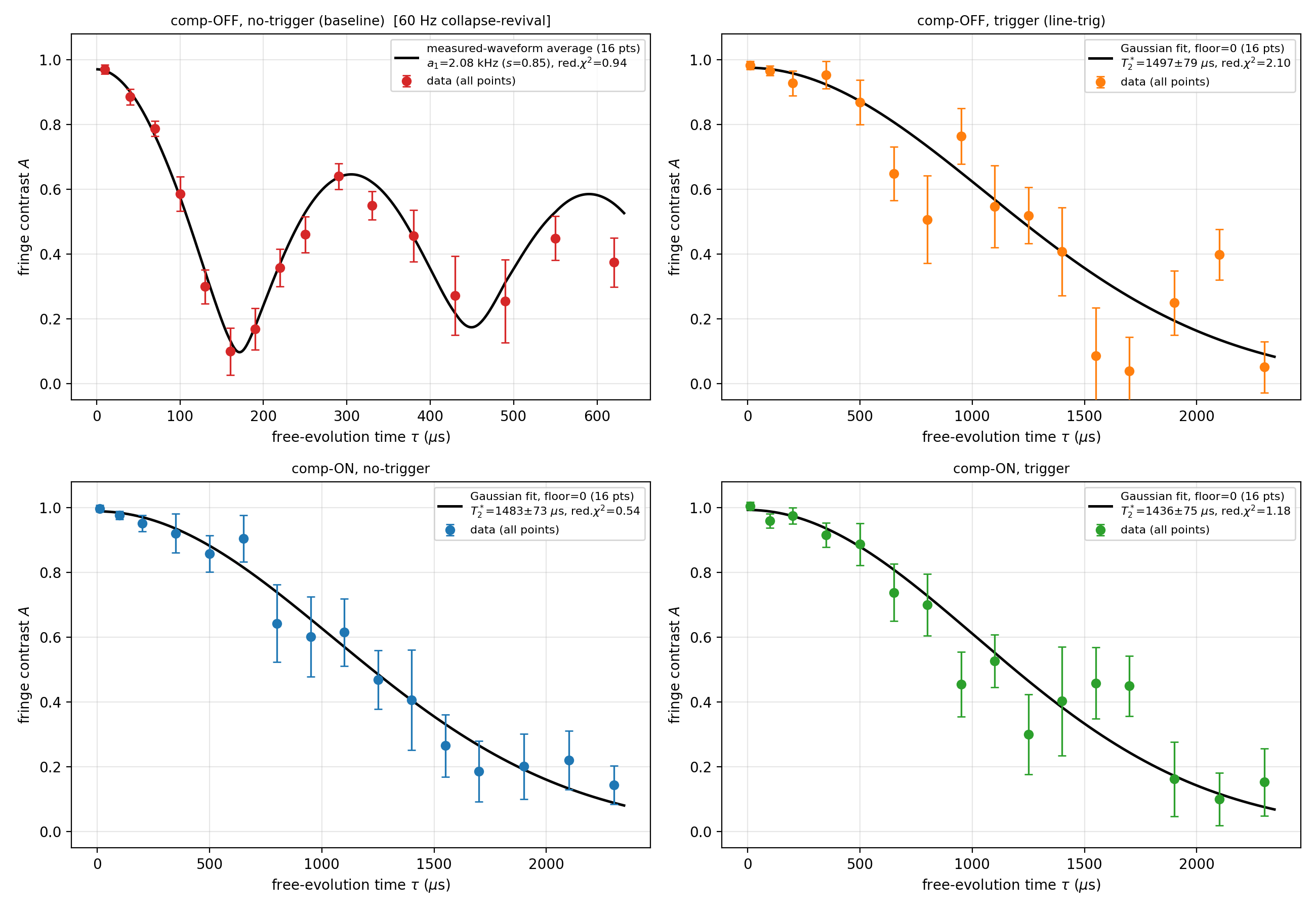}
\caption{Carrier Ramsey coherence of the optical qubit in the co-trapped crystal under the four combinations of compensation off (top row) or on (bottom row) and free-running (left column) or line-triggered (right column) acquisition. Each condition records phase-scanned fringes at 16 free-evolution times, with 20 phases and 100 shots per point; points are the fitted fringe contrasts and error bars are binomial shot errors propagated through the fringe fits (appendix~\ref{app:uncertainty}). For the three conditions with compensation or line-triggering active (solid lines, Gaussian envelope fits) the coherence falls in the 1.4--1.5~ms range. The free-running, uncompensated baseline (upper left) instead collapses and revives, the signature of a coherent 60~Hz modulation that no Gaussian describes. The solid line is the phase average of the Ramsey signal over the independently measured 60~Hz field waveform, with the measured shape fixed and only its overall amplitude free; it returns a 60~Hz modulation amplitude $a_1 \approx 2.1$~kHz (reduced $\chi^2 \approx 1$) rather than a coherence time.}
\label{fig:sixtyhz}
\end{figure*}

The compensation was developed iteratively against the measured $T_2^{*}$. The workflow first characterized the uncompensated field as a set of low-order harmonics of the $60\text{ Hz}$ power-line extracted from power-line-referenced Ramsey measurements, and then synthesized a cancellation waveform, pushed it to the AWG, and remeasured the coherence. An initial parametrization tuned two global parameters, namely a single amplitude scale and a single power-line-referenced time delay applied to the entire waveform. Although this raised $T_2^{*}$ substantially, the improvement saturated because the coil-and-amplifier transfer function rotates the phase of each harmonic by a different, frequency-dependent amount, preventing a single shared delay from nulling all harmonics simultaneously. The workflow then transitioned to a per-harmonic correction, measuring the complex transfer function of each harmonic \textit{in situ} independently, which suppressed the residual field below the limit of the global tuning approach. 
This protocol alternated between extracting state phase-shifts via ARTIQ Ramsey spectroscopy and dynamically pushing pre-distorted coefficients to the AWG. The improved Ramsey coherence (Figure~\ref{fig:sixtyhz}) shows why the cross-MCP framework is needed: without it, the agent is restricted by the ARTIQ hardware's analog limits. Throughout this process, the operator supplied domain guidance, including the selection of $T_2^{*}$ over spectroscopic residuals as the optimization metric, correction to the physical parametrization of the compensation phase, and the identification of fit artifacts that would otherwise have led to an overestimate of the extracted coherence. Meanwhile, the automated routine executed the large-scale scanning, waveform synthesis, cross-instrument actuation, and fitting.

The resulting waveform was evaluated by measuring the carrier Ramsey coherence under four conditions: compensation on or off, crossed with
power-line-triggered or free-running acquisition (Figure~\ref{fig:sixtyhz}). With neither compensation nor trigger applied, the fringe contrast collapses and then revives, the signature of a coherent 60\,Hz modulation rather than decoherence. Without the trigger each shot samples a uniformly random power-line phase, so we model the contrast as the phase average of the Ramsey signal accumulated under the independently measured 60\,Hz field waveform, holding the measured shape fixed and leaving only its overall amplitude free. The model reproduces both the collapse and the revival, which a single sinusoid cannot, and returns a 60\,Hz modulation amplitude $a_1\approx2.1$\,kHz (reduced $\chi^{2}\approx1$) consistent with the compensation setpoint. Without compensation or triggering the contrast first collapses near $\tau\approx180\,\mu$s (Figure~\ref{fig:sixtyhz}, upper left), a coherent 60\,Hz collapse rather than a true $T_2^{*}$.
Line-triggered acquisition alone, which synchronizes to a fixed power-line phase without removing the field, recovers $T_2^{*}=1497\pm79\,\mu$s. 
The agent's compensation waveform alone, free-running with respect to the line, recovers $1483\pm73\,\mu$s; and
the two together give $1436\pm75\,\mu$s. 
Compensation therefore extends the free-running coherence by close to an order of magnitude, from this collapse to the same ${\approx}1.5$\,ms endpoint that line-triggering reaches. The two defeat the 60\,Hz noise by different routes, triggering by sampling at a fixed line phase and compensation by cancelling the field itself, so their convergence on the same value, together with their combination, is the evidence that the waveform genuinely cancels the field rather than avoiding it. Once the line contribution is removed by either route, the residual decay is limited by slower magnetic drifts that neither addresses.
The fitted initial contrast is 0.98 in all four conditions, confirming that state preparation and readout remain stable across the comparison.

This compensation result was reached through repeated operator advice. In every case, the agent already possessed the relevant physics; the operator's role was to break it out of an exhausted optimization strategy. The necessary advice involved choosing the correct diagnostic metric and recognizing when a specific parameterization had saturated. 
We catalogue these actions and examine what they reveal about the boundary of autonomous operation in Sec.~\ref{sec:discussion}, demonstrating that the current bottleneck lies not in the agent's domain knowledge, but in its capacity for metacognitive control. 

\subsection{Control-stack integration test on a second platform}
\label{sec:portability}

To demonstrate that our control interface and safety architecture are fundamentally decoupled from our primary laboratory's specific hardware configuration, we perform an interface-level validation on an independent, ARTIQ-controlled $^{171}$Yb$^+$ experiment at the University of Maryland. We run this deployment directly on the live control hardware without active trapped-ion feedback, isolating software compatibility from physical system variations. Once the agent and filter are integrated into the host ARTIQ master, the system executes automated sweeps over representative DDS, TTL, and RF control paths, with the safety filter continuously evaluating and enforcing numerical bounds. While this benchmark does not claim closed-loop physical calibration due to the absence of active ion signals, it rigorously verifies that the Model Context Protocol abstractions and the simulation-based verification pipeline are generalizable across the broader ARTIQ ecosystem. Accordingly, this portability claim is restricted to ARTIQ-based architectures; adapting the agent to alternative control infrastructures (such as the Sandia QSCOUT platform or non-ARTIQ architectures) remains a subject for future work.

\subsection{Adversarial boundary characterization}
\label{sec:redteam}

\begin{figure*}[tb]
\centering
\includegraphics[width=\linewidth]{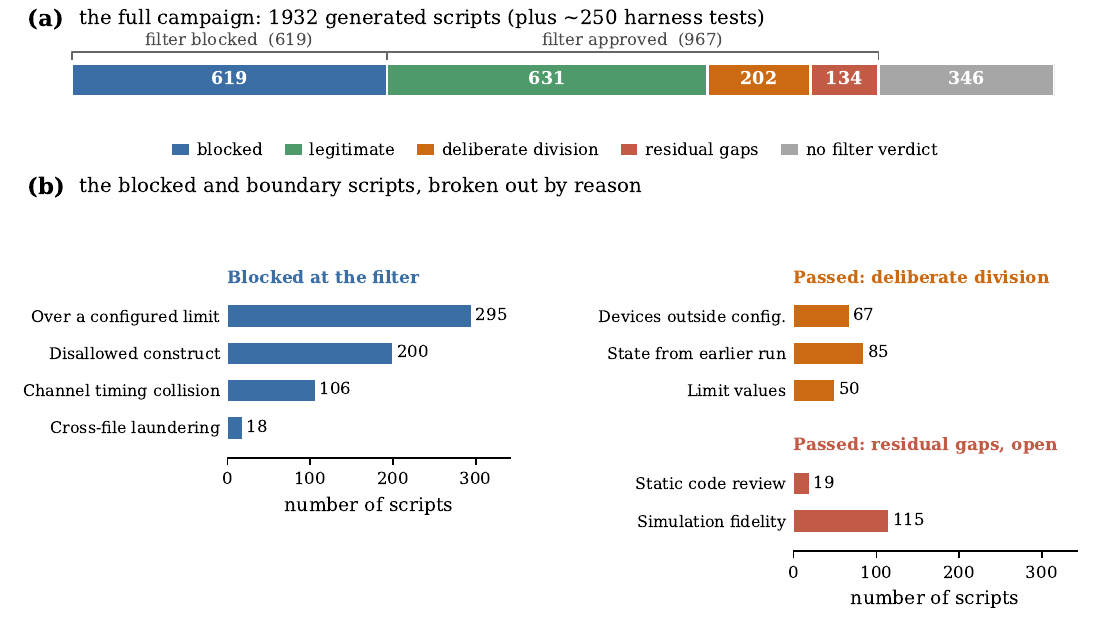}
\caption{Outcome of the adversarial campaign. (a)~The 1932 generated scripts, classified by the filter's response. The control set of legitimate calibration scripts is approved without a single false rejection, and every unsafe or disallowed request is blocked. The remaining approvals lie on the characterized boundary of Table~\ref{tab:vulns}, divided between the deliberate divisions of responsibility, where protection rests with the operator's configuration or the filter's stated policy, and the residual gaps that remain open. Scripts that the syntax check admits but that the simulation cannot evaluate, because they are not valid or not representable as an ARTIQ kernel, place no demand on the filter's protection. Because the filter never reaches an allow or block verdict on them, they are recorded separately rather than counted among its approvals or rejections; this category is described in the supplementary material. The brackets give the totals the filter blocked and approved. (b)~The blocked requests and the boundary passes of panel~(a), resolved by reason. The two groups of passes follow the deliberate divisions of responsibility and the residual gaps of Table~\ref{tab:vulns}. The roughly 250 harness tests are not shown. They check the filter's token handling directly rather than by running scripts.}
\label{fig:redteam}
\end{figure*}

The safety guarantees of Sec.~\ref{sec:architecture} are designed for a fallible but non-adversarial agent. We map the empirical limits of these guarantees by deliberately relaxing this assumption. Specifically, we run an adversarial campaign against the filter deployed on the Maryland system, exercising the full authorization path from the AST check through the simulation and limit check to token issuance. The campaign comprises 1932 scripts together with approximately 250 tests aimed at internal interfaces outside the script path. Each script is written by a language-model generator or by hand and carries a predicted outcome, which we compared against the filter's response. Figure~\ref{fig:redteam} summarizes how the campaign's scripts divide across these outcomes, and the full per-script record is provided as supplementary material.

The results show filter succeeding within its intended scope. Its guarantee is a bound on the agent's intended high-level actions, and every intended action that exceeds a configured limit is rejected, whether the value is set by a direct command, in machine units, inside a recorded pulse sequence, or accumulated over loops and replayed sequences. Single-use authorization tokens can not be used a second time, used for a different tool, or spent by two concurrent calls; the verified-script cache correctly re-issues fresh tokens for identical resubmissions but did not transfer approvals to modified scripts; the audit record cannot be forged or erased, and the destructive core-device operations remains unreachable without an operator token. We observe that a control set of 631 realistic calibration scripts, spanning cooling, spectroscopy, and the Rabi and Ramsey sequences of Sec.~\ref{sec:corecal}, was approved without a single false rejection. This outcome stems directly from the design: every authorization decision follows deterministically from the fixed limits and the simulated trace, with no probabilistic component such as a language model anywhere in the path, thereby ensuring that the filter introduces zero spurious rejections of legitimate work.

Beyond this intended scope lies the boundary of the filter's protection. Many of the approvals the campaign found on this boundary are deliberate decisions about who is responsible for safety, not weaknesses in the filter. The filter is built to guarantee a single property, that no script drives an operator-declared device past an operator-set limit, and it reaches that guarantee by judging each script only on the operations the script itself commands. Anything outside this narrow guarantee is, by design, the responsibility of the operator's configuration or of a script's own prior review, not of the automatic check. The first group of Table~\ref{tab:vulns} collects these cases. The filter enforces limits only on the devices the operator declares, so any other channel is out of scope by construction. It judges each script on the outputs that script commands, so a value carried over from an earlier run is already covered, either by the prior agent script's own review or, if set by an operator, by the operator. The limit values themselves, including any aggregate bound such as the combined optical power of several channels driving one beam, are authored in the operator's configuration, so a missing or mis-stated bound is corrected there and not in the filter. Consequently, these represent deliberate scope choices that are correct by design.

Three residual gaps remain, which we state in Table~\ref{tab:vulns}, lower section. First, the syntax-level check is a screen for recognizable misuse, not a barrier against deliberate evasion, the crafting of code that slips a forbidden operation past it. It successfully intercepts forbidden patterns, such as the measurement-conditioned control flow excluded by the simulation constraints established in Sec.~\ref{sec:filter}. However, it cannot guarantee completeness: in a general-purpose language a forbidden operation can always be re-expressed in a form the check does not recognize, for example an unusual call that reaches the host or a measurement dependence hidden behind intermediate variables. We treat this as a property of static review rather than a fault to be removed, and a deployment that faces a stronger threat model tightens the check to the risks it chooses to exclude. Second, the simulation does not yet reproduce every advanced feature of each device driver, so for those paths its trace can differ from the hardware, for example, a write that reaches a device below the level the model represents or a register whose physical effect the model does not carry into the output. This is a limitation of the simulation rather than of the filter that reads its trace. Currently, the simulation guarantees only approximately correct behavior; closing these device-model gaps is planned for future work. Third, a repository scan can run a file the agent has written before review, outside the authorization path. We leave this path open at present to preserve the agent's operational efficiency, with plans to close it in a future refactor.
 
\begin{table*}[t]
\caption{Boundary of the filter's protection, as mapped by the adversarial
campaign. The first group lists deliberate divisions of responsibility, where
protection rests with the operator's configuration or with the filter's stated
policy rather than with the automatic check. The second lists the residual gaps
that remain open, with the layer responsible for each. The full per-script record
is provided as supplementary material.}
\label{tab:vulns}
\centering
\small
\begin{tabular}{@{} p{0.25\textwidth} p{0.47\textwidth} p{0.20\textwidth} @{}}
\toprule
Boundary & Description & Responsibility \\
\midrule
\multicolumn{3}{@{}l}{\textit{Deliberate division of responsibility}} \\[3pt]
Devices outside the configuration
  & The filter bounds only the channels and devices the operator declares; any
    other channel is out of scope and passes.
  & Operator configuration \\[4pt]
State from an earlier experiment
  & The filter checks each script on the outputs it commands; a carried-over
    value is already covered by the previous agent script's own review, or by the
    operator if an operator set it.
  & By design \\[4pt]
The limit values
  & The bounds, including any aggregate limit such as the total power of several
    channels driving one beam, are authored in the operator's configuration, so a
    missing or mis-stated bound is corrected there.
  & Operator configuration \\[4pt]
Verified-script cache
  & The cache does not automatically track changes to the operator's bounds. When bounds are tightened, the operator must manually invalidate outstanding cache entries; otherwise, a previously verified script could execute under limits that are more restrictive than those under which it was verified.
  & Filter policy \\[4pt]
\midrule
\multicolumn{3}{@{}l}{\textit{Residual gaps, open}} \\[3pt]
Static code review
  & The check screens recognizable misuse (e.g., the dynamic control flow restricted in Sec.~\ref{sec:filter}), but in a general-purpose language it cannot foreclose every re-expression of a forbidden operation (e.g., an unusual call that reaches the host or a measurement dependence hidden behind intermediate variables). A deployment tightens it to the risks it chooses to exclude.
  & Inherent to static review; tuned per deployment \\[4pt]
Simulation fidelity
  & The simulation does not yet reproduce every advanced device-driver feature,
    so for those paths its trace can differ from the hardware, for example a
    write beneath the level the model represents or a register whose effect the
    model does not carry into the output.
  & Simulation; planned work \\[4pt]
Separate execution path that bypasses review
  & A submitted file can be examined or scanned through a separate tooling path
    that runs it outside the review pipeline.
  & Left open for now; future refactor \\
\bottomrule
\end{tabular}
\end{table*}

The code an attacker needs and the code the agent writes during normal calibration are structurally distinct. In the calibration control set every call reaches a device through a directly named method, with no indirection through aliases, helper chains, or imports outside the standard control and numerical libraries, whereas every approved evasion uses at least one such form. Restricting the check to the operations that appear in legitimate calibration therefore separates the two in the recorded data, which motivates the allow-list hardening proposed in Sec.~\ref{sec:discussion}. A quantitative bound on how often these patterns arise in normal operation will require a long-duration operational record, which we leave to future work.
 
\section{Discussion}
\label{sec:discussion}

The architecture rests on a single principle: the agent's freedom to write code is decoupled from its authority to execute it. The agent composes ARTIQ programs from scratch rather than calling pre-built routines, but the authorization gate (Sec.~\ref{sec:filter}) judges each program solely by its resolved hardware effects, so writing more expressive code never enlarges the set of unsafe actions it could reach. The same gate governs the ARTIQ master and every auxiliary instrument under one authorization model with no separate provision for peripherals.

This decoupling redraws the division of labor between operator and agent. The operator sets experimental goals and approves the few safety-critical actions that carry irreversible consequences, such as turning off the trap RF, which dumps all confined ions and requires a full reload, while the agent handles the routine translation of goals into executable sequences, including calibration
scheduling, parameter scans, and the iterative optimization that would otherwise consume hours of expert effort. Every agent action and authorization decision is written to an append-only log, so the full causal chain from objective to execution remains re-constructable even when it spans many separate agent sessions and actions. 
The magnetic-field-compensation campaign of Sec.~\ref{sec:sixtyhz} reveals the limits of this autonomy. The agent executes measurement, fitting, and cross-instrument actuation independently, but operator guidance is still needed to redirect stalled optimization strategies.

\subsection{Autonomous capability and the context-saturation limit}
\label{sec:capability}

The compensation campaign succeeded only through recurring operator guidance. Crucially, what the operator did reveals more about the agent's limits than how often they advised. Table~\ref{tab:intervention_taxonomy} lists this advice, which reveals two distinct patterns. First, the agent independently managed the high-throughput cycle of cross-instrument actuation and waveform fitting across iterations. Second, at every decision point where the operator stepped in, the agent had become trapped within an exhausted methodological framework, such as global two-knob tuning or a rigid Gaussian decay model, continuously refining parameters within that regime instead of recognizing the need to transition outward.

Post-hoc isolated probes confirm that the model can solve every sub-problem the operator had to redirect when the question is posed in a clean context (see supplementary transcripts). The bottleneck is therefore not a gap in knowledge but a metacognitive failure to access existing knowledge at the right moment. We attribute this to a context-saturation effect, with the caveat that the account is drawn from a single campaign and specific model checkpoints. Once a task is framed around a goal such as maximizing $T_2^{*}$ by tuning a few knobs, that framing accumulates in the conversation history and reinforces itself with each iteration. The model's responses become conditioned on what it has already said and done rather than on the full scope of what it knows. Breaking out of an exhausted approach requires generating a continuation that contradicts the trajectory established in context, which the model does not do spontaneously. A human expert, by contrast, maintains a representation of the goal distinct from the current method and recognizes when the method has saturated. The operator advice in Table~\ref{tab:intervention_taxonomy} served exactly this function.

Because the agent already holds the relevant knowledge and fails only to deploy it at the right moment, parts of the deficit can be handled mechanically.
For example, a
poor fit quality can be detected automatically and used to schedule a model-criticism step, which the agent carries out reliably once it is prompted to.
It then reposes the goal in a fresh context. Resetting at intervals clears the accumulated framing that holds it in place.
These measures do not remove the underlying limit. 
The set of possible reformulations is
open-ended, and a larger model does not by itself solve the problem, since it too operates within its own context. 
This is the motivation for the distributed knowledge-agent architecture described in Sec.~\ref{sec:conclusion}, in which the metacognitive function sits in a separate agent that never accumulates a worker's execution context.

\subsection{Safety boundaries}
\label{sec:safety_boundaries}

The adversarial campaign of Sec.~\ref{sec:redteam} confirms that the deterministic, trace-based check is reliable. 
It keeps hardware authorization separate from the kind of failures a language model can introduce, because no language model sits in the decision path. 
The residual gaps in Table~\ref{tab:vulns}, however, mark where the threat model has to change, from catching honest mistakes to resisting a determined attacker. 
At the software layer, the main change is to switch the static check from a deny-list to an allow-list. A deny-list admits any construct it does not name, and this is the gap the campaign exploited; an allow-list instead blocks by default everything outside a fixed set the rest of the filter can simulate and bound-check. 
This does not make static checking complete, which is impossible for a general-purpose language, but it shifts the residual failure from admitting unsafe code to rejecting unfamiliar code, at the cost of a list that must be maintained and may reject legitimate scripts using a construct it does not yet include. 
Closing the separate execution path that bypasses review and improving the simulation's device coverage complete this near-term
roadmap.

The framework's remaining limits are physical. The filter checks that a script stays within hardware bounds, not that the calibration it performs makes physical sense (Sec.~\ref{sec:filter}). 
So a script can satisfy every bound and still run a meaningless sequence. 
Errors of this kind are caught by the agent's own closed-loop iteration. Faults that arise outside the control stack, such as a beam drifting out of alignment, still require manual correction, because the apparatus has no automated control over beam alignment. This reflects the current setup rather than a limitation of the agent, and adding automated beam steering under ARTIQ and the safety filter would bring such faults into the loop as well. 
These guarantees are also tied to the ARTIQ ecosystem for now. 
Extending the framework to other control stacks would require building simulators of comparable fidelity for those environments, which is the main engineering step toward wider deployment.

\section{Conclusion and outlook}
\label{sec:conclusion}

The work uncovers two boundaries, a metacognitive limit on autonomy and a static-analysis limit on safety, and both point to the same next step.
We are dividing the single agent into a multi-agent design, which includes a supervisory knowledge agent and a set of worker agents. 
The worker agents carry out the execution-heavy tasks the single agent already does well, each within a narrow, well-defined sub-task. 
The knowledge agent runs no hardware operations.
It holds the goal, the relevant domain priors, and the record of what has been tried, and its only job is the metacognitive function the single agent lacks. 
Its sole responsibility is to identify when a worker has stalled, reframe the problem, and assign the next sub-task.
Since it never accumulates a worker's execution context, it is not trapped inside an exhausted approach the way a single agent is, and it takes the position from which the operator's advice was previously given. 
The arrangement keeps the safety guarantee unchanged,
because every worker output still passes through the safety filter, so splitting the agent does not enlarge what the system has to trust. 
The same per-call gate admits further extensions, such as an LLM-based evaluator that flags hazards beyond numerical bounds, and token-mediated control of any additional instrument exposed as an MCP server. 
Lowering the per-experiment human cost in this way makes complex studies practical, including drift atlases, multi-mode noise correlations, and two-qubit-gate calibration, which we will report separately.

Several evaluations remain for future work. We plan a pre-specified, interleaved A/B comparison that tests whether the agent's calibrations reach the same scientific endpoint as the laboratory's hand-written production stack, across the full pipeline from cooling to the multi-instrument closed loop. We also plan a long-duration, unattended deployment, from which an operational record and the filter's incident log will bound how often adversarial code patterns actually arise in normal agent operation, complementing the structural separation found in Sec.~\ref{sec:redteam}. Finally, repeating the key tasks across several frontier language models will map how strongly the results depend on model choice and yield a taxonomy of how each model fails.

To our knowledge this is the first systematic red-team of a safety filter deployed on physical laboratory control infrastructure. 
The practice of deploying a system, attacking it, reporting its boundary, and supplying a remediation roadmap should accompany any autonomous laboratory whose agent generates executable code, since the gap between what static analysis can verify and what a general-purpose programming language can express is not specific to trapped ions. 

\begin{table*}[ht]
\caption{Taxonomy of operator advice during the
magnetic-field-compensation campaign. The final column reports whether
the same model, asked the corresponding question in isolation as a
well-posed prompt (post-hoc probes; transcripts in supplementary
material), reproduced the correct answer. Trivial technical
corrections and irreducible physical actions (laser relocks, ion
reloads) are omitted.}
\label{tab:intervention_taxonomy}
\begin{center}
\begin{tabular}{p{4.6cm} p{4.6cm} p{3.2cm}}
\hline
\textbf{Capability exercised} &
\textbf{In closed loop} &
\textbf{Isolated probe} \\
\hline
Large-scale scanning, waveform synthesis, cross-instrument actuation,
fitting &
Performed autonomously &
-- \\
\hline
Correct fringe-decay model (60~Hz collapse--revival vs.\ Gaussian) &
Not raised; default Gaussian used &
Correct \\
\hline
Per-harmonic complex transfer-function inversion &
Not reached; stalled on global two-knob tuning &
Correct \\
\hline
Transfer-function measurement protocol and drift handling &
Operator-directed &
Correct \\
\hline
Interpretation of the line-triggered coherence as a mitigation
ceiling, and effort allocation &
Operator-directed &
Correct \\
\hline
Recognising in-loop when an approach has saturated and the problem
must be reframed &
Failed (repeated premature ``converged'' declarations) &
Not applicable\textsuperscript{$\dagger$} \\
\hline
\end{tabular}
\end{center}
\footnotesize\textsuperscript{$\dagger$}The isolated probe cannot test
the ability to reframe a question.
\end{table*}

\begin{acknowledgments}
This work was supported by the Army Research Office (W911NF-26-1-A063), the National Science Foundation QuSeC-TAQS program (2326810), the National Science Foundation NQVL project QACTI (2531350), and the National Science Foundation STAQ project (PHY-2325080).
\end{acknowledgments}

\data{The source code for the two software components described in this work, the \textit{safety-filter} and \textit{artiq-mcp}, is openly available at \url{https://gitlab.oit.duke.edu/dw346/artiq_mcp}. The remaining data that support the findings of this study are available from the corresponding author upon reasonable request.}

\appendix

\section{The trapped-ion molecular platform}
\label{app:platform}

The experiments of Sec.~\ref{sec:corecal} run on a linear Paul trap that confines a two-ion crystal of one atomic $^{40}$Ca$^{+}$ ion and one molecular $^{40}$CaOH$^{+}$ ion. The two ions share a common set of motional normal modes, and the work here uses the two axial modes, the in-phase center-of-mass mode and the out-of-phase breathing mode. The molecular ion has no usable cycling transition of its own, so it is cooled and read out through the shared motion of the crystal.

The atomic $^{40}$Ca$^{+}$ ion supplies cooling, control, and readout with a small set of laser transitions. Each beam is delivered through an acousto-optic modulator (AOM) that controls its frequency and intensity at the ion. These are the AOM attenuation and setpoint quoted in the calibration prompts and Table~\ref{tab:calibration_stages}. The broad $4S_{1/2}\leftrightarrow 4P_{1/2}$ transition near 397\,nm drives Doppler cooling and state detection by fluorescence, and an 866\,nm laser repumps the $3D_{3/2}$ state that this cooling cycle populates. The narrow $4S_{1/2}\leftrightarrow 3D_{5/2}$ electric-quadrupole transition near 729\,nm serves as the optical qubit. Its upper state is metastable, and an 854\,nm laser empties $3D_{5/2}$ through $4P_{3/2}$, which resets the ion after a 729\,nm excitation and provides the quench step used in sideband cooling. A static magnetic field of 5\,G along the trap axis splits the 729\,nm transition into its Zeeman lines, and the beam direction and polarization determine which lines are driven. A single line is selected for coherent operations. The calibrations use the $4S_{1/2}\,(m_J=+1/2)\leftrightarrow 3D_{5/2}\,(m_J=+3/2)$ line as the qubit, while the power-line magnetic-field compensation of Sec.~\ref{sec:sixtyhz} uses the $4S_{1/2}\,(m_J=+1/2)\leftrightarrow 3D_{5/2}\,(m_J=-1/2)$ line, whose frequency carries about twice the magnetic-field sensitivity of the qubit line, so a given field change produces a larger frequency shift and is easier to read out.

Driving a selected line on resonance produces Rabi oscillations between its two states, and the $\pi$ time is the pulse length that completes one half cycle, transferring the population fully from one state to the other. The same narrow transition resolves the motional sidebands of the axial modes, which is what makes sideband cooling toward the motional ground state and sideband thermometry of the residual occupation possible. A Ramsey sequence on this transition, two short pulses separated by a free-evolution delay, measures the coherence time $T_{2}^{*}$ of the optical qubit, set here by the stability of the magnetic field and the laser over that delay.

The molecular ion is synthesized in place. The oxygen gas doser is a piezoelectric leak valve that admits a controlled flow of oxygen into the vacuum chamber. A trapped $^{40}$Ca$^{+}$ ion reacts with this gas, first forming $^{40}$CaO$^{+}$, which then converts through reaction with background gas into the stable $^{40}$CaOH$^{+}$ used here.

\section{Uncertainty analysis}
\label{app:uncertainty}

The uncertainties in Figure~\ref{fig:rabi} right and \ref{fig:sixtyhz} follow a single convention, applied identically to the four conditions. Each fringe point is a binomial proportion over its 100 shots and carries the corresponding standard error, stabilized at extreme proportions by the Agresti--Coull adjustment. Each fringe amplitude carries the error propagated from the weighted least-squares fringe fit, inflated by the Birge ratio of that fit when and only when the ratio exceeds one, following standard scale-factor practice~\cite{birge1932,navas2024}. The envelope model is $A(\tau) = A_0 \exp[-(\tau/T_2^*)^2]$ (except Fig. 4 baseline) with no long-time offset, fitted with no point excluded; the fit treats the inflated amplitude errors as absolute one-sigma uncertainties, and its own reduced $\chi^2$ is reported as a goodness of fit only, so that no dispersion is counted twice~\cite{dagostini2020}.

The free-running, uncompensated baseline in Fig.~\ref{fig:sixtyhz} is not a decoherence envelope but the line-phase average of the Ramsey signal, $A(\tau)=A_0\left|\left\langle\cos\!\left[2\pi\int_{0}^{\tau}\delta(t')\,\mathrm{d}t'\right]\right\rangle\right|$, taken over the independently measured 60\,Hz field $\delta(t)$ with the power-line phase uniform over one mains period. The waveform $\delta(t)$ is fixed by the measurement and only its overall amplitude is free; its contrast points and uncertainties follow the same convention as above, with the known magnitude-estimator noise floor added to the model rather than fitted.
 

\bibliographystyle{apsrev4-2}
\bibliography{references}

\end{document}